\documentclass[aps,prb,twocolumn,floatfix,groupedaddress,showkeys]{revtex4}
\usepackage{graphicx}
\begin{document}

%\preprint{}

\title{Angular Dependence of Switching Properties in Single Fe Nanopillars}


\author{G.~Brown}\email[]{browngrg@csit.fsu.edu}
\altaffiliation[Also at]
{
CSIT,
Florida State University, Tallahassee, FL 32306, USA}
\affiliation{Center for Computational Sciences, 
Oak Ridge National Laboratory, Oak Ridge, TN 37831, USA}
\author{S.M.\ Stinnett}
\author{M.A.\ Novotny}\email[]{man40@ra.msstate.edu}
\affiliation{Department of Physics and Astronomy and ERC Center for 
Computational 
Sciences, Mississippi State University, Mississippi State, MS 39759, USA}
\author{P.A.\ Rikvold}\email[]{rikvold@csit.fsu.edu}
\affiliation{Department of Physics, Center for Materials Research and 
Technology, and School for Computational Science and Information Technology, 
Florida State University, Tallahassee, FL 32306-4350, USA}

\date{\today}

\begin{abstract}
The continued increase in areal densities in magnetic recording makes it
crucial to understand magnetization reversal in nanoparticles. We present
finite-temperature micromagnetic simulations of hysteresis in Fe
nanopillars with the long axis tilted at angles from 0 degrees to 90 degrees 
to the applied sinusoidal field. The field period is 15 ns, and the particle 
size is $9 \times 9 \times 150$ nm.  The system is discretized into a 
rectangular pillar of $7 \times 7 \times 101$ spins each with
uniform magnetization. At low angles, reversal begins at the endcaps 
and proceeds toward the center of the particle. At ninety degrees, reversal 
proceeds along the entire length of the particle 
(save at the ends).  The switching field was observed to increase
over the entire range of angles, consistent with
recent experimental observations. A second, lower-resolution micromagnetic 
simulation with $1 \times 1 \times 17$ spins, does not agree with 
experiment, but shows behavior very similar to that
of the Stoner-Wohlfarth model of coherent rotation.
\end{abstract}


\maketitle

\begin{figure}[tb]
\includegraphics[angle=-90,width=.47\textwidth]{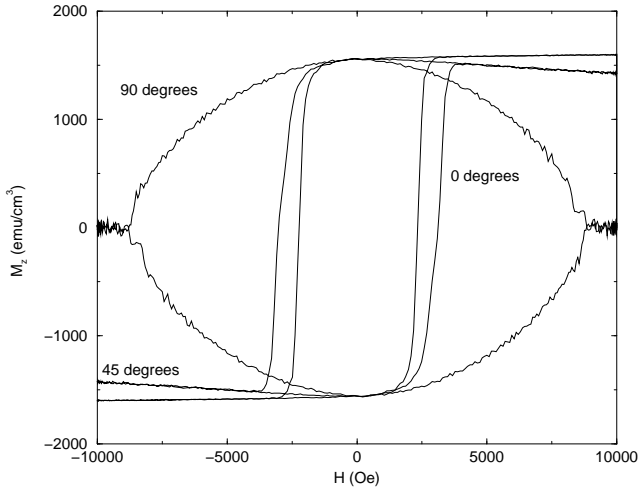}
\caption[]{
Hysteresis Loops for the $4949$-spin model with 
misalignment angles of $0^\circ$, $45^\circ$, 
and $90^\circ$.  The applied field is sinusoidal with a period of 15~ns, 
and the temperature is $T=100$~K. 
One half of a loop was calculated and the data reflected 
to show a complete loop.}
\end{figure}

Hysteresis in magnetic particles has been studied
for nearly a century. However, the increasing areal densities currently 
being pursued in magnetic recording technologies are pushing bit sizes
to length scales where the atomic nature of materials is important.
In addition, large data-transfer rates
require switching times not much longer than the precession times
of atomic moments. Taken together, further technological advances require
a better understanding of hysteresis in nanoscale magnets at nanosecond 
time scales.
Since these nanomagnets are often single-domain, the cooperative effects
among the atomic spins determine the switching behavior.
This cooperative behavior can be studied with simulations of 
realistic model systems, and can be directly
compared to high-resolution experimental results,\cite{WIRTH,WIRTH99,LI2,LI} 
possibly for individual nanomagnets. \cite{LI}
In the present paper we present such dynamic simulations of hysteresis 
in nanometer-sized Fe nanopillars exposed to a magnetic field that is 
aligned at an angle with the pillars' long axis. 

The general method in micromagnetics simulations
is to divide a model magnetic system into 
coarse-grained regions, each with a uniform magnetization density
$\vec{M}(\vec{r}_i)$ of a constant magnitude, $M_S$.
Here $\vec{r}_i$ is the position of the $i$th spin in the 
system.  The time evolution of each spin is given by the 
Landau-Lifshitz-Gilbert (LLG) equation,\cite{brown,gregbrown1,Nowak}
\begin{equation}
\frac{d \vec{M}(\vec{r}_i)}{d t}
=
\frac{\gamma_0}{1+\alpha^2}(\vec{M}(\vec{r}_i)
\times[\vec{H}(\vec{r}_i)-\frac{\alpha}{M_S}(\vec{M}(\vec{r}_i)
\times\vec{H}(\vec{r}_i))])
\end{equation}
where $\vec{H}(\vec{r}_i)$ is the total local field at the $i$th position, 
$\gamma_0$ is the gyromagnetic ratio ($1.76\times10^7$~rad/Oe~s), 
and $\alpha = 0.1$ is a dimensionless damping parameter which determines 
the rate of energy dissipation.  For the sign of the precession term, 
we follow the convention of Brown.\cite{brown}

The total local field, $\vec{H}(\vec{r}_i)$, includes contributions 
from the applied field, dipole-dipole fields, and exchange interactions;
\cite{gregbrown1} here the small crystalline anisotropy of Fe is ignored.  
At nonzero temperatures, thermal fluctuations also contribute a term in 
the form of a stochastic thermal field which fluctuates independently for 
each spin.  The thermal fluctuations are Gaussian with a mean of zero 
and covariance given by the fluctuation-dissipation 
theorem.\cite{gregbrown1}

We have used two different 
models to simulate hysteresis in single Fe nanopillars, as have
been synthesized and studied by Wirth, {\em et al}. \cite{WIRTH,WIRTH99,LI2,LI}
The first model is a 9~nm$\times$9~nm$\times$150~nm pillar,  
discretized into a cubic lattice with $4949$ sites (7$\times$7$\times$101).
For such a large number of sites,
direct calculation of the dipole-dipole interactions is impractical,  
and the Fast Multipole Method was employed, as described
elsewhere.\cite{gregbrown1} 
The second model is a 5.2~nm$\times$5.2~nm$\times$88~nm Fe
pillar discretized into a one-dimensional chain of 17 spins. 
This model assumes that 
deviations of the magnetization 
from uniformity over the cross-section of the pillar are 
negligible, as has been observed in Monte Carlo simulations. \cite{Nowak}
For both models, the material properties were chosen to correspond to bulk Fe,  
with saturation magnetization $M_{\rm S}$$=$$1700$~emu/cm$^3$ and 
exchange length 2.6~nm. 

\begin{figure}[tb]
\includegraphics[angle=0,width=.47\textwidth]{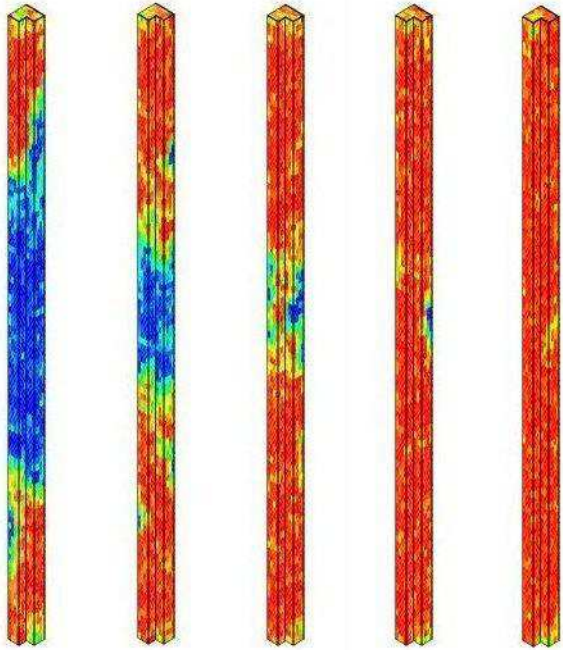}
\caption[]{
The $z$-component of the magnetization at times $t$$=$
$4.48$, $4.53$, $4.58$, $4.63$, and $4.68$ ns 
during the switching process of the $45^\circ$ hysteresis loop 
in Fig.~1. 
}
\end{figure}

\begin{figure}[tb]
\includegraphics[angle=0,width=.47\textwidth]{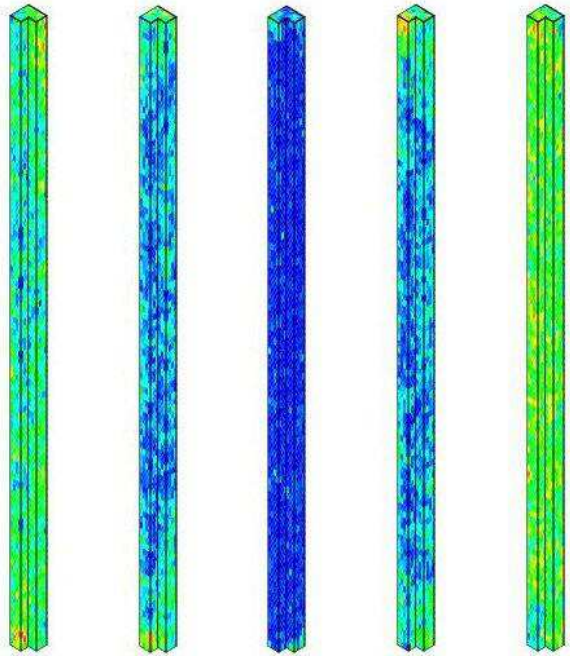}
\caption[]{
The $z$-component of the magnetization at times 
$t$$=$ $2.25$, $3.75$, $5.25$, $6.00$, and $6.75$ ns
during the relaxation process of the $90^\circ$ hysteresis loop 
in Fig.~1. 
}
\end{figure}

\begin{figure}[tb]
\includegraphics[angle=-90,width=.47\textwidth]{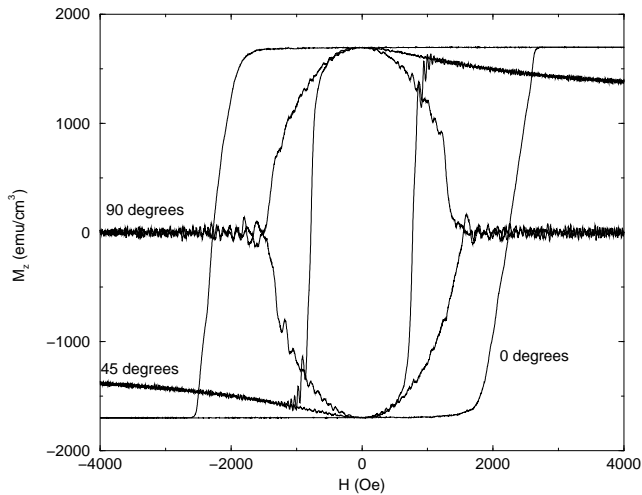}
\caption[]{
Hysteresis loops at $0^\circ$, $45^\circ$, and $90^\circ$ 
misalignment between the field and easy axis for the $17$-spin
model.  The applied field is sinusoidal with period 200~ns, and 
$T=10$~K. Qualitatively, the loop shapes appear similar to those 
for the $4949$-spin model (Fig.~1),  but the angular-dependence of the
switching field for the two models is quite different.
}
\end{figure}

\begin{figure}[tb]
\includegraphics[angle=0,width=.47\textwidth]{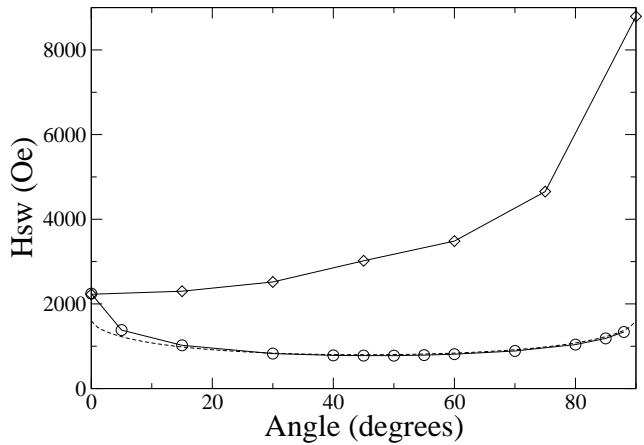}
\caption[]{
The angular dependence of the switching field $H_{\rm sw}$, shown 
the $4949$-spin model (diamonds) and the $17$-spin model (circles).
The shape for the $17$-spin model is qualitatively 
similar to what is expected from the Stoner-Wohlfarth model (dashed curve), 
while the $4949$-spin model yields a dependence more consistent with recent 
experiments.
}
\end{figure}

Figure~1 shows hysteresis loops at $T= 100$~K for the first model with the 
field misaligned 
at $0^\circ$, $45^\circ$, and $90^\circ$ to the long axis of the 
pillar.  The loops were calculated using a sinusoidal field with a period 
of 15~ns, which started at a maximum value of 10,000~Oe.  
In all the loops shown here, the reported magnetization is the component 
along the long axis ($z$-axis) of the pillar. With the field and pillar aligned
($0^\circ$), reversal initiates at the ends, as previously reported. 
\cite{gregbrown1} At $45^\circ$ misalignment between pillar and field, 
the magnetization is initially pulled away from the long (easy) axis by the 
large magnetic field.  As the field is swept toward zero, the magnetization 
relaxes until it essentially reaches saturation at zero applied field.

Figure~2 shows the $z$-component of the magnetization at selected times 
during the reversal process for the $45^\circ$ hysteresis loop in Fig.~1.  
As with $0^\circ$,  regions of reversed magnetization nucleate at the ends of 
the pillar, and it is the growth of these reversed regions that leads 
to magnetic reversal of the particle.  It is important to note that 
the particles do not have a uniform magnetization, even though they 
are single-domain particles and would be in equilibrium under quiescent
conditions.

For $90^\circ$ misalignment, the reversal mechanism is quite different.  
The hysteresis loop in Fig.~1 
shows that the magnetization is essentially perpendicular to the easy 
direction until the field reaches a particular value.  As the field is 
decreased further, the magnetization relaxes toward the easy axis.  
Since nothing breaks the up/down symmetry of the system when the applied field
has no component along the easy axis, the relaxed magnetization can be 
directed toward either the positive or negative $z$-axis. 
Figure~3 shows the $z$-component of the magnetization for the $90^\circ$ 
misalignment at selected times during the 
hysteresis loop. For this case, the relaxation occurs along the entire length
of the pillar, except at the ends. The pole-avoidance effect present at the ends,
and involved in nucleation at smaller angles, retards relaxation along the easy axis.

The hysteresis loops for the second, smaller model shown in Fig.~4 are 
qualitatively similar to those of Fig.~1.  Again loops at $0^\circ$, 
$80^\circ$, and $90^\circ$ misalignment are shown.  There are important
differences between the two models, however.  First, without 
lateral resolution of the magnetization across the cross-section, these 
pillars exhibit ringing due to the processional dynamics.
Evidently, the precession of individual moments in the $4949$-spin model does 
not lead to precession of the end-cap moment; possibly the spin waves 
rapidly damp out the gyromagnetic motion.  A second, 
and more prominent, difference between the models is observed in the angular
dependence of the switching field, $H_{\rm sw}$, shown in Fig.~5.  
Here $H_{\rm sw}$ is defined as the applied field at 
which $M_z = 0$ (or in the case of $90^\circ$ misalignment, where 
$M_z \neq 0$).  The second model (circles) shows a shape qualitatively 
similar to what is expected from Stoner-Wohlfarth (SW) theory,
with a minimum $H_{\rm sw}$ at $45^\circ.$
The dashed curve is the SW theory with $H_k$$=$$1600$ Oe, much smaller than
the $10^4$ Oe assumed for ellipsoids. The first model (diamonds), on the 
other hand, has minimum $H_{\rm sw}$ at $0^\circ,$ and increases as 
the misalignment angle is increased. This behavior is consistent with 
recent experimental observations of Fe nanopillars.\cite{WIRTH,LI,LI2}

In summary, we have presented results from large-scale, finite-temperature 
micromagnetics simulations of hysteresis in models of nanometer-sized Fe 
pillars in magnetic fields inclined with respect to the long axis 
of the pillars. In a model in which the magnetization 
is constant across the short dimension of the pillar, the angular 
dependence of the switching field is similar to that expected from 
Stoner-Wohlfarth theory. However, in a model with spatial resolution 
across the short dimension, the switching field has a monotonically 
increasing angular behavior consistent with recent experimental results.

%\begin{acknowledgments}
We acknowledge support from NSF grant No.\ DMR-0120310 and the DOE Office
of Science through the Computational Material Science Network of BES-DMSE.
%\end{acknowledgments}


\begin{thebibliography}{4}

\bibitem{WIRTH}
S.~Wirth, M.~Field, D.D.\ Awschalom, and S.\ von~Moln{\'a}r, 
Phys.\ Rev.\ B {\bf 57}, R14028 (1998). 

\bibitem{WIRTH99}
S.~Wirth and S.~von~Moln{\'a}r,
J.\ Appl.\ Phys. {\bf 85}, 5249 (1999).

\bibitem{LI2}
Y.~Li, P.~Xiong, S.~von~Moln{\'a}r, Y.~Ohno, and H.~Ohno,
Appl.\ Phys.\ Lett.\ {\bf 80}, 4644 (2002).

\bibitem{LI}
Y.~Li, P.~Xiong, S.~von~Moln{\'a}r, Y.~Ohno, and H.~Ohno,
J.\ Appl.\ Phys.\ {\bf 93}, 7912 (2003).

\bibitem{brown}
W.~Brown, {\it Micromagnetics} (Wiley, New York, 1963).

\bibitem{gregbrown1}
G.Brown, M.A.\ Novotny and P.A.\ Rikvold,
Phys.\ Rev.\ B {\bf 64}, 134432 (2001).

\bibitem{Nowak}
U.~Nowak, in
{\it Annual Reviews of Computational Physics IX},
edited by D.~Stauffer (World Scientific, Singapore, 2001), p.~105.

\end{thebibliography}
\end{document}